\documentclass[aps,prl,twocolumn,letterpaper,floatfix,showpacs,superscriptaddress]{revtex4-1}
\usepackage{graphicx}
\usepackage{amsmath,amssymb,amsfonts} 
\usepackage{mathtools}
\usepackage[squaren]{SIunits}
\usepackage[hidelinks]{hyperref}
\usepackage{epstopdf}
\usepackage{color}

\newcommand{\fpf}[1]{({\MakeLowercase{#1}})}
\newcommand{\fpl}[1]{({\MakeLowercase{#1}})}
\newcommand{\fpt}[1]{\MakeLowercase{#1}}

\newcommand{\coloronlineornothing}{(Color online) }

\definecolor{mymagenta}{rgb}{1,0,1}
\definecolor{mycyan}{rgb}{0,1,1}
\definecolor{mydarkgreen}{rgb}{0,0.5,0}
\usepackage{tikz}
\usetikzlibrary{plotmarks}
\newcommand\marksymbol[4]{\tikz[#2,scale=#4,rotate=#3]\pgfuseplotmark{#1};}
\newcommand{\blackdash}{\textcolor{black}{--}}
\newcommand{\cyandash}{\textcolor{mycyan}{--}}
\newcommand{\reddash}{\textcolor{red}{--}}
\newcommand{\trianglelo}{\protect\marksymbol{triangle}{mymagenta}{90}{1.4}}
\newcommand{\trianglelf}{\protect\marksymbol{triangle*}{mymagenta}{90}{1.4}}
\newcommand{\diamondo}{\protect\marksymbol{diamond}{black}{0}{1.4}}
\newcommand{\diamondf}{\protect\marksymbol{diamond*}{black}{0}{1.4}}
\newcommand{\circleo}{\protect\marksymbol{o}{blue}{0}{1}}
\newcommand{\circlef}{\protect\marksymbol{*}{blue}{0}{1}}
\newcommand{\squareo}{\protect\marksymbol{square}{mydarkgreen}{0}{1}}
\newcommand{\squaref}{\protect\marksymbol{square*}{mydarkgreen}{0}{1}}
\newcommand{\triangledo}{\protect\marksymbol{triangle}{red}{180}{1.4}}
\newcommand{\triangledf}{\protect\marksymbol{triangle*}{red}{180}{1.4}}

\newcommand{\Fig}{Figure}
\newcommand{\fig}{Fig.}

\providecommand{\abs}[1]{\lvert#1\rvert}
\newcommand{\Vfluid}{V_\text{fluid}}
\newcommand{\Vfluidtransition}{\Vfluid^\text{transition}}
\newcommand{\Vobject}{\dot X}
\newcommand{\Vobjectaverage}{\langle\Vobject\rangle}
\newcommand{\Vstrongzero}{V_0^\text{strong}}
\newcommand{\Vweakzero}{V_0^\text{weak}}
\newcommand{\Fb}{F_b}
\newcommand{\koff}{k^\text{off}}
\newcommand{\koffzero}{k^\text{off}_0}
\newcommand{\kon}{k^\text{on}}
\newcommand{\konzero}{k^\text{on}_0}

\newcommand{\Ntype}{N_\text{type}}
\newcommand{\Ntypes}{N_\text{types}}


\begin{document}
\title{Dynamical bond cooperativity enables very fast and strong binding between sliding surfaces}
\author{J{\o}rgen Kjoshagen Tr{\o}mborg}
\email{j.k.tromborg@fys.uio.no}
\affiliation{Department of Physics, University of Oslo, Sem S{\ae}lands vei 24, NO-0316, Oslo, Norway}
\affiliation{Massachusetts Institute of Technology, 77 Massachusetts Avenue, Cambridge, MA 02139, USA}
\author{Alfredo Alexander-Katz}
\email{aalexand@mit.edu}
\affiliation{Massachusetts Institute of Technology, 77 Massachusetts Avenue, Cambridge, MA 02139, USA}
\date{\today} 

\begin{abstract}
Cooperative binding affects many processes in biology, but it is commonly addressed only in equilibrium. In this work we explore dynamical cooperativity in driven systems, where the cooperation occurs because some of the bonds change the dynamical response of the system to a regime where the other bonds become active. To investigate such cooperativity we study the frictional binding between two flow driven surfaces that interact through a large population of activated bonds. In particular, we study systems where each bond can have two different modes: one mode corresponds to a fast forming yet weak bond, and the other is a strong yet slow forming bond. We find considerable cooperativity between both types of bonds. Under some conditions the system behaves as if there were only one binding mode, corresponding to a strong and fast forming bond. Our results may have important implications on the friction and adhesion between sliding surfaces containing complementary binding motifs, such as in the case of cells binding to the vessel walls under strong flowing conditions.
\end{abstract}

\maketitle
Many biological functions depend on the ability to arrest objects traveling in the blood stream by binding them to the vessel wall. Examples include the binding of platelets to stem hemorrhage \citep{Jackson2007growing,Springer2014von} and the binding of immune cells to the vessel wall prior to their entry into adjacent tissue \citep{Springer1994traffic,Harlan1985leukocyte-endothelial}. These bonds are formed between dedicated proteins existing in the blood stream, on the platelets or immune cells, or on the vessel wall. In order to arrest the flowing cells, several ligand-receptor bonds need to form, particularly at high flow rates. 
Such ligand-receptor bonds, as for example selectin and its ligand in leukocytes \citep{Springer1994traffic}, or glycoprotein GpIb$\alpha$ in platelets and the A1 domain in Von Willebrand Factor (VWF) \citep{Sadler1998biochemistry, Springer2014von, Sing2013von}, are protein complexes that contain elementary bonding units such as salt bridges and hydrogen bonds, as well as hydrophobic interactions.
To simultaneously activate multiple elementary bonds or effective hydrophobic interactions requires configurational registry between the protein and ligand, which means that the strongest bonds tend also to be the slowest. 

Interestingly, recent developments have shown exceedingly rapid adhesion of platelets to VWF; the time scales are the fastest that have been reported for ligand-receptor interactions \citep{Wellings2012mechanisms, Bark2012correlation,Nesbitt2009shear}. This behavior is still far from understood, yet it is critical for the formation of plugs \citep{Springer2014von, Chen2013blood-clotting-inspired, Schneider2007shear-induced, Alexander-Katz2014toward}. Here, we demonstrate that a system in which we combine multiple types of bonds can exhibit such behavior. In particular, we study a simple model of how an object embedded in a fluid and flowing over a substrate arrests under the action of a cascade of binding events. We are interested in a scenario in which a single kind of bond is not sufficient to arrest the system. The first kind of bond is not strong enough, the second is not fast enough. In this scenario we study the effect of dynamical cooperativity between the bonds when both types are present, and find the conditions under which they together confer their favorable properties (being fast or being strong) on the system. Our results indicate that depending on the degeneracy of the bonds and the dynamical properties of each, it is possible to create a system which appears as if it contains a single ``hybrid bond'' type that is fast forming yet strong. This is quite relevant in fast sliding conditions such as blood flow, where common biological bonds may be too slow to significantly affect the dynamics. Such a scenario is particularly important during uncontrolled formation of platelet-VWF aggregates in stenotic regions that can lead to stroke or heart infarcts \citep{Sanders2013reduced, deMeyer2012von,Nesbitt2009shear} . 

Our work sheds light on the importance of weak bonds in out-of-equilibrium conditions. The existence of weak bonds has long been speculated to have very important implications in binding because they may increase the apparent on-rate substantially. For example, it was recently pointed out that a variety of binding sites can be present on protein surfaces, and that even weaker binding regions can significantly affect the kinetics of binding \citep{Baron2013molecular,Kim2008replica,Schilder2013formation,Shan2011how}.


\begin{figure}
\centering
\includegraphics[width=\columnwidth]{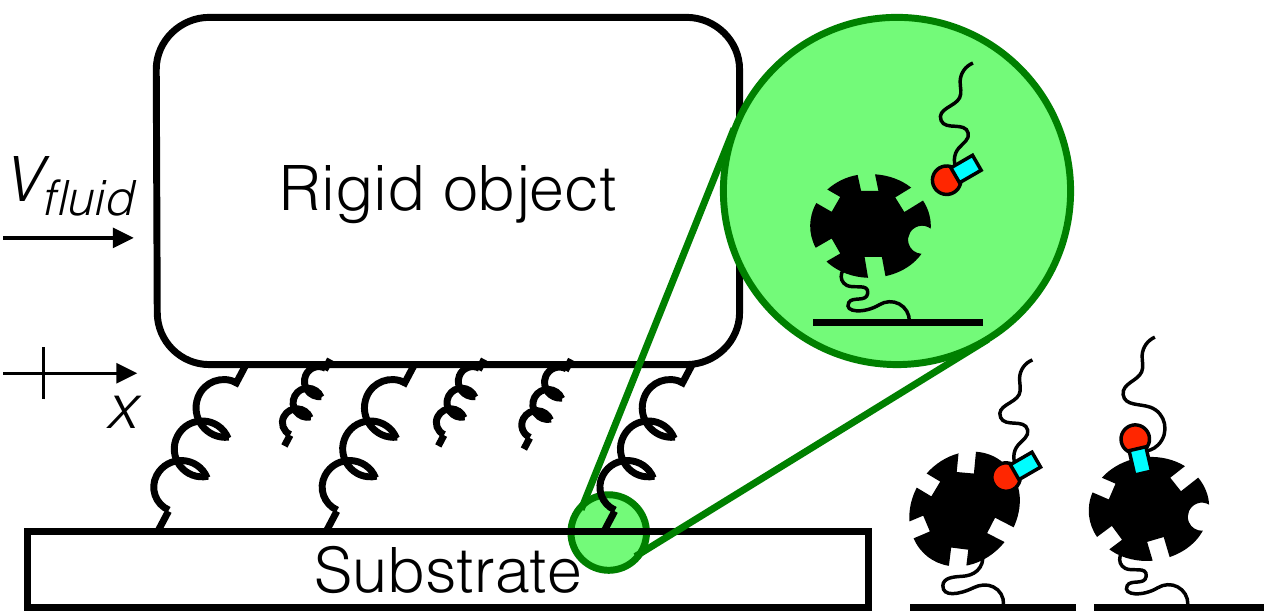}%
\caption{\coloronlineornothing Sketch of the model system. A rigid object interacts with the substrate though the formation and breaking of bonds. Each bond can be in a broken, a strong or a number of weak configurations.\label{fig:sketch}}
\end{figure}

Our model is related to the method called adhesive dynamics \citep{Hammer1992simulation,Chang2000state,King2001multiparticle,Korn2008dynamic, Mody2008platelet}, but we have simplified the flow problem in order to focus solely on the interplay between the bond types, which to the best of our knowledge has not been addressed before. The model is sketched in \fig~\ref{fig:sketch}. It consists of a substrate and a rigid object immersed in the fluid flow above the substrate, and the bonds between them. The substrate can for example represent the wall of a blood vessel and the rigid object a platelet in the blood stream. The bonds between the substrate and the rigid object are modeled as springs. The intact bonds contribute a friction force that resists the motion of the object, slowing it down or arresting it depending on the bond properties and flow conditions. The equation of motion of the rigid object is
\begin{align}
M\ddot X = \eta(\Vfluid-\dot X) + \Fb,
\end{align}
where $M$ is the object's mass, $\ddot X$ is its acceleration (the dots denote time derivatives), $\eta$ is the effective viscosity of the object in the fluid, $\Vfluid$ is the speed of the fluid with respect to the substrate, $\dot X$ is the object's instantaneous speed with respect to the substrate and $\Fb$ is the force from the intact bonds,
\begin{align}
\Fb = \sum_i^N f_i^{(x)}.
\end{align}
$N$ is the number of bonds that can form. Broken bonds contribute zero force. The force in an intact bond is $f_i = \kappa(l_i-l_0)$, where $\kappa$ is the bond stiffness, $l_i$ is the length of the bond and $l_0$ is an equilibrium bond length; $l_i=\sqrt{x_i^2+h^2}$ where $x_i$ is the elongation of the bond along the $x$ direction and $h$ is the vertical separation of the object and the substrate. The projection of $f_i$ along the direction of the fluid flow is $f_i^{(x)} = -f_ix_i/l_i$.
When a bond is intact the elongation of the bond occurs with the same speed as the motion of the object, $\dot x_i=\dot X$. When the bond breaks, its elongation returns to zero, controlled by the parameter $\lambda$ as follows
\begin{align}
\dot x_i =
\left\{
\begin{array}{ll}
	\dot X &, \text{intact},\\
	-\lambda x_i &, \text{broken}.
\end{array}
\right.
\end{align}

The rupture and formation of bonds is controlled by two rates that we label $\koff$ and $\kon$. These depend on the type of bond being broken or formed (weak or strong). In each time step $\Delta t$ of our simulation an intact bond breaks with probability $\koff\Delta t$. We draw a random number from the uniform distribution $\xi\in[0,\,1]$ and break the bond if $\xi<\koff\Delta t$. For bond formation, we first draw a random number $\xi$ to determine which type of bond to attempt to form, with each bond type weighted by their relative abundance. Thus, if there are 5 different weak bonds and 1 strong bond, $\Ntype=[5,\,1]$, the attempt is for a weak bond if $\xi<5/(5+1)$ and for a strong bond otherwise. This can be generalized to any number of bond types. Then, another $\xi$ is drawn and the bond forms if $\xi<\Ntypes\kon\Delta t$. Multiplying by the sum of bond types, $\Ntypes=\sum_i N_{\text{type},i}$, ensures that the bond formation attempt probabilities are additive, so that the probability to form each bond is independent of the number of bond types being considered. The rates $\kon$ are thus characteristic of each type of bond, while the number of potential bonds of each type is encoded in $\Ntype$. Even though bond formation attempts are additive, however, each bond can only be of one type at any given time (or the bond can be broken).

For the bond rupture rates we use the simple Bell-type expression \citep{Bell1978models}
\begin{align}
\koff = \koffzero \exp\left(\beta f_i\Delta x\right),\label{eq:koff}
\end{align}
where $\koffzero$ is the bond rupture rate in the absence of any force on the bond, $\beta=1/(k_BT)$ and $\Delta x$ is the length to the transition state from the bottom of the well. For biological protein-protein complexes this scale is on the order of a nanometer. \Citet{Filippov2004friction}, studying a similar model, introduce another rupture rate expression for what they call strong bonds, but we have used equation~\eqref{eq:koff} for both types of bond, assigning different $\koffzero$ to weak and strong bonds.

The rate of bond formation when the object is stationary with respect to the substrate is $\konzero$ for each type of bond. The rate of bond formation when the object is in motion is the product of the stationary rate and a function $g()$ that reduces the rate. This function effectively introduces a limiting timescale for the formation of such bonds. This is natural in all systems, as there is always a relaxation timescale upon which certain binding configurations can be reached. One can consider this timescale as an effective attempt rate. In the case of proteins, one expects such a time scale to be in the millisecond range. Interestingly, in the case of the interaction between GpIb$\alpha$ and A1 such interaction appears to occur within microseconds. When two surfaces slide past each other,  the time available to form the bond is $\tau=a/\abs{\dot X}$, where $a$ is the lateral extent of the reaction site. We assign a characteristic time $\tau_0$ (or equivalently a characteristic velocity $V_0$) for each bond type, $\tau_0=a/V_0$. The full rate of bond formation is
\begin{align}
\kon = \konzero g\left(\frac{\tau-\tau_0}{\Delta\tau}\right),
\end{align}
with $\Delta\tau$ a parameter that controls the sharpness of the reduction in $\kon$ with $\abs{\dot X}$. We have used $g(\frac{\tau-\tau_0}{\Delta\tau}) = \tanh(\frac{\tau-\tau_0}{\Delta\tau})$.
We express length, time, and force in units of $h$, $M/\eta$, and $k_BT/h$, respectively.

We set as initial conditions that the object is moving with the fluid with no intact bonds, $\dot X=\Vfluid$, $l_i = l_0\forall i$. We then integrate the motion of the rigid object forwards in time using the leap-frog method with time step $\Delta t$. The resulting motion has three distinct regimes: (i) the object moves with the flow with very little friction from the substrate and the average speed of the object $\Vobjectaverage\approx\Vfluid$, (ii) significant friction from the substrate slows the object down to $\Vobjectaverage$ markedly smaller than $\Vfluid$, and (iii) the object comes nearly to rest. The regime selected depends on the properties of the bonds available and on the flow speed $\Vfluid$.

\begin{figure}
\centering
\includegraphics[width=\columnwidth]{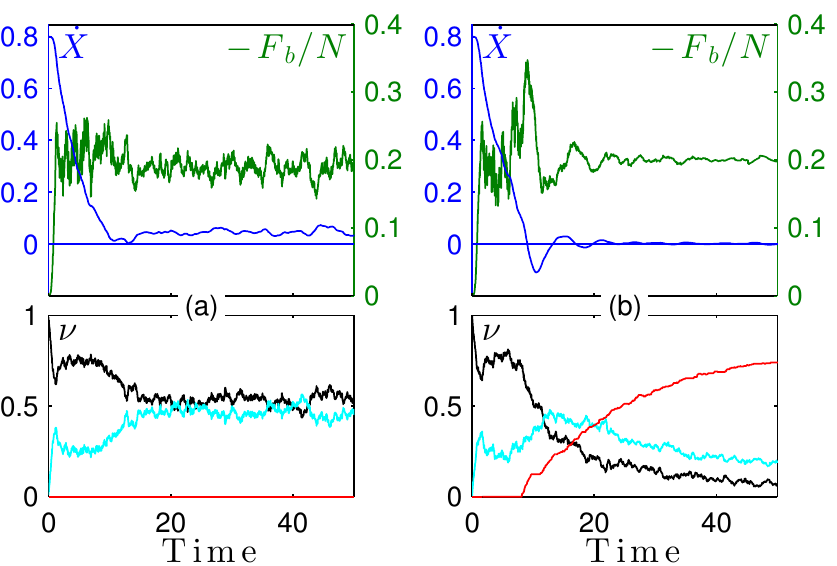}%
\caption{\coloronlineornothing Example time evolution of object speed and bond properties when \fpf{a} only weak (but fast) bonds are possible, and when \fpf{b} both weak (but fast) and strong (but slow) bonds are possible. Top panels: object speed and average force in bonds. Bottom panels: fractions $\nu$ of unformed (\blackdash), weak (\cyandash) and strong (\reddash) bonds. In this paper $N=300$, $\eta=N/4$, $M=N$, $\kappa=[10,\, 10]$ (for $[\text{weak},\,\text{strong}]$ bonds), $l_0=h=1$, $\lambda=0.4\eta/M$, $\koffzero=[0.2,\,0.001]$, $\konzero=[0.1,\,0.1]$, $V_0=[\Vweakzero,\,\Vstrongzero]=[1,\,0.1]$, $a=[1,\,1]$, $\Delta\tau=[0.1,\,0.1]$, $\beta=1$, $\Delta x=[1,\,1]$, $\Delta t=0.01$. In this figure $\Ntype=[5,\,0]$ \fpl{a} or $\Ntype=[5,\,1]$ \fpl{b}, $\Vfluid=0.8$.\label{fig:time_evolution}}
\end{figure}

\Fig~\ref{fig:time_evolution}\fpt{a} shows a typical evolution for weak bonds. The population of bonds quickly stabilizes around an equilibrium value, with bonds being broken and formed continuously. The object slows down to an average speed $\Vobjectaverage\approx 0.1$, markedly smaller than $\Vfluid=0.8$, but far from coming to rest. \Fig~\ref{fig:time_evolution}\fpt{b} shows a typical evolution when a strong bond option is added; the settings are otherwise the same as in \fig~\ref{fig:time_evolution}\fpt{a}. The beginning of the dynamics is the same. The weak bonds form and the speed of the object is reduced. Notice that for the first $18$ or so units of time, the number of strong bonds remains negligible. Then, as the object's speed falls to $\Vobject\approx\Vstrongzero$ and the probability to form strong bonds increases significantly, the strong bonds take over and the object's speed eventually falls to $\Vobject\approx 0$. For additional data on steady-state bond fractions see Supplemental Material at the end of this file.

\begin{figure}
\centering
\includegraphics{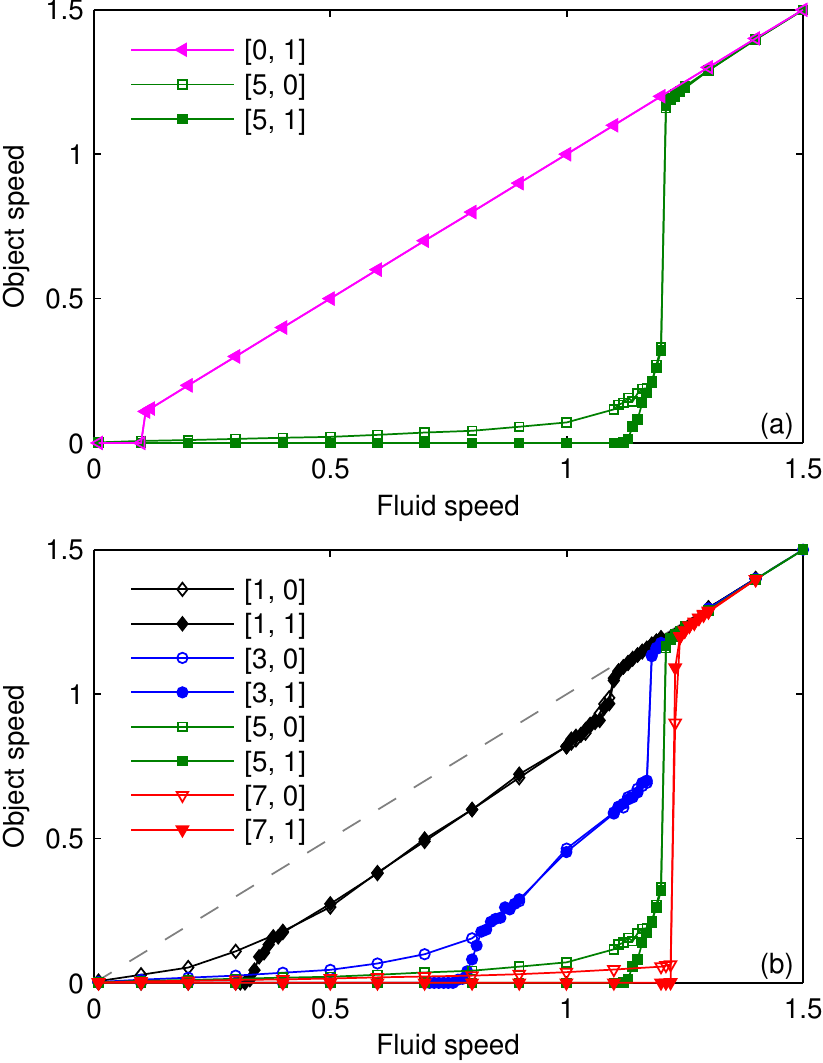}%
\caption{\coloronlineornothing Average steady-state speed of rigid object vs fluid flow speed $\Vfluid$ for systems with different bonds available. \fpf{a} Strong only, weak only, and a combination. \fpf{b} Weak only and the corresponding weak plus one strong bonds. The gray dashed line shows $\Vobject=\Vfluid$ as a guide to the eye. Legends show values of $\Ntype$. $\Vobjectaverage$ was measured over the second half of simulations lasting for $150$ units of time. Points near the transition from arrest to non-arrest are averages over ca. 20 simulations.\label{fig:Vslider_vs_Vfluid}} 
\end{figure}

\Fig~\ref{fig:Vslider_vs_Vfluid} shows the average object speed $\Vobjectaverage$ as a function of fluid flow speed $\Vfluid$ for a range of scenarios for what bonds are available. \Fig~\ref{fig:Vslider_vs_Vfluid}\fpt{a} shows three cases. (i) If only strong (slow) bonds exist, the object moves as fast as the fluid for all but the lowest fluid speeds. For $\Vfluid<\Vstrongzero$ strong bonds are formed and bring the object to rest. (ii) If only weak (fast) bonds exist, the object is slowed down by the weak bonds for $\Vfluid\lessapprox\Vweakzero$, but the bonds are unable to bring the object to rest. (iii) If weak (fast) and strong (slow) bonds both exist, their combined effect brings the object to rest for $\Vfluid\lessapprox\Vweakzero$. \Fig~\ref{fig:Vslider_vs_Vfluid}\fpt{b} shows the combination of 1, 3, 5, or 7 weak bonds with 0 or 1 strong bond. When there are fewer weak bonds they are less effective at slowing the object down, and the transition to arrest in the weak plus strong case moves to lower values of $\Vfluid$. Note that for $\Ntype=[7,\,1]$ the combined dynamics is equivalent to what we would expect from having bonds that are strong \emph{and} fast. Thus, we can say that the fast bonds lend their speed to the slow bonds, or equivalently that the strong bonds lend their strength to the weak bonds. This synergistic effect is what we termed dynamical cooperativity. 

\begin{figure}
\centering
\includegraphics{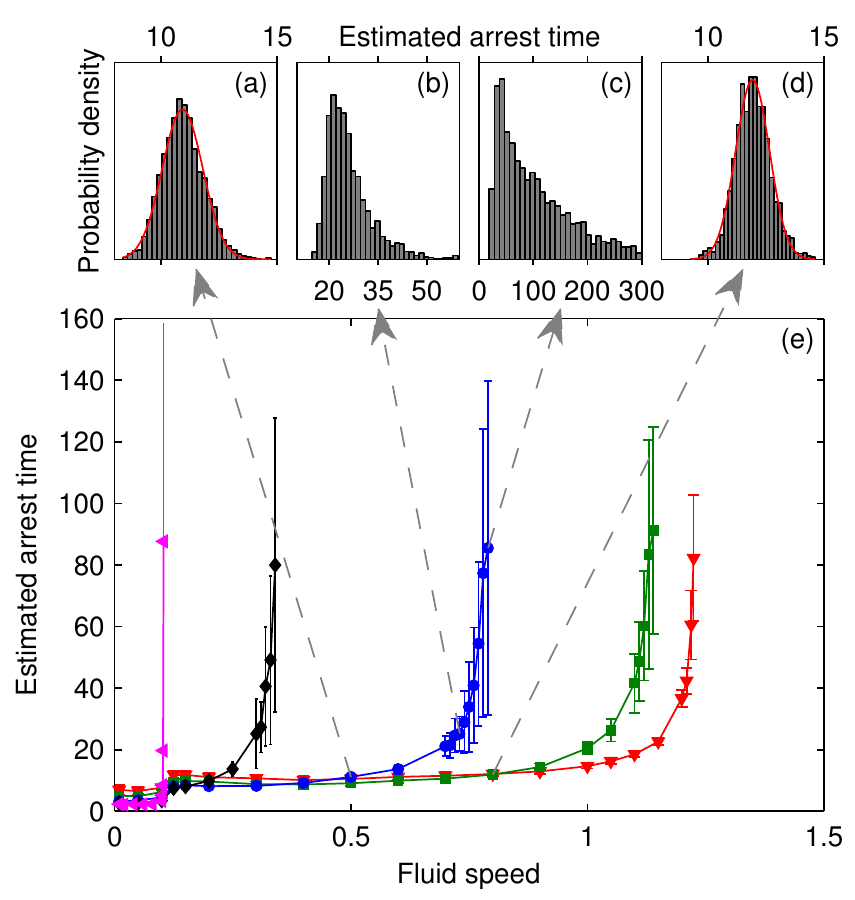}%
\caption{\coloronlineornothing Estimated object arrest time. \fpf{a--d} Distributions of arrest times for selected simulation settings (arrows originate on the $\Ntype=[3,\,1]$ and $[5,\,1]$ lines). The fits in \fpl{a} and \fpl{d} are normal distributions. Each histogram is based on ca. $2000$ simulations. \fpf{e} Estimated arrest time vs $\Vfluid$. Colors and markers are the same as in \fig~\ref{fig:Vslider_vs_Vfluid}. For the points close to where each line diverges, the total simulation time was increased to $300$ units of time. Error bars show one standard deviation away from the mean. Away from $\Vfluidtransition$ the error bars are smaller than the markers. Each point is based on 40 simulations.\label{fig:arrest_time}}
\end{figure}

For the conditions that lead to the object coming to rest, \fig~\ref{fig:arrest_time}\fpt{e} shows the arrest time as a function of $\Vfluid$. In all our simulation the arrest occurs after the fraction of strong bonds starts growing. We use the first time when $\nu_\text{strong bonds}>0.2$ as an estimate for the arrest time, because this is numerically more robust than a test on $\Vobject$, which can fluctuate around or near zero, see \fig~\ref{fig:time_evolution}. For each combination of available bonds there is a transition speed $\Vfluidtransition$ above which strong bonds do not form. The arrest time diverges as $\Vfluid\rightarrow\Vfluidtransition$ from below. Away from the divergence, there is only a weak dependence of the arrest time on $\Vfluid$. This can be understood from \fig~\ref{fig:Vslider_vs_Vfluid}\fpt{b}. When the weak bonds alone lead to an average speed $\Vobjectaverage<\Vstrongzero$ the object first slows down under the influence of the weak bonds, then the strong bonds take over and the object arrests. This process depends only weakly on $\Vfluid$ as long as $\Vfluid<\Vweakzero$. When the weak bonds alone lead to $\Vobjectaverage\gtrapprox\Vstrongzero$, however, the strong bonds only form when fluctuations in the object's speed momentarily bring $\Vobject$ below $\Vstrongzero$. Thus the arrest time becomes dependent on the extreme value statistics of the fluctuations in $\Vobject$. \Fig~\ref{fig:arrest_time}\fpt{a--d} show the distribution of arrest times for selected conditions. Away from $\Vfluidtransition$ (\fpt{a,d}) the arrest times are well fitted by normal distributions. Closer to $\Vfluidtransition$ (\fpt{b,c}) the distributions of arrest times change and become strongly asymmetric.

In this work we have demonstrated that the kinetics of binding is strongly modified when one has dynamical cooperativity in which two different types of bonds exchange their dynamical properties. In our case we showed that under some conditions one can obtain complete ``hybridization'' of the properties, and the system appears as having bonds that are fast \emph{and} strong, although one starts from a mixture of slow and strong bonds together with weak and fast bonds. Such cooperativity is interesting because it relies heavily on the degeneracy of the weak bonds, and not on the strong bonds. In other words, it is dominated by the kinetics of the bonds, and not the equilibrium properties. 


In conclusion, for strongly out-of-equilibrium situations we have found that the presence of the weak bonds is crucial, and enables the system to achieve states (such as the arrested state) that would otherwise be unattainable. This leads us to suggest that experimental observations of apparently fast and strong bonds may in fact be due to dynamical cooperativity rather than to a single bond type. The question is how to detect the presence of the weak bonds. In principle, this could be possible by studying the rolling of beads under an applied torque as a function of the rotational frequency \cite{Steimel2014artificial}. If the timescales are well separated, one should see two different regimes of friction dominated by the different bond types. Clearly, such dynamical cooperativity would be of much importance in situations where one would like to ``recycle'' different binding motifs, and not synthesize de novo a special binding partner with the prescribed kinetic attributes. Thus, we expect this behavior to be important in many diverse areas of lubrication, friction and adhesion. Furthermore, confirming and characterizing such weak interactions could lead to progress in our understanding of the equilibrium binding properties of proteins \citep{Schilder2013formation,Baron2013molecular, Kim2008replica, Shan2011how}. 

\begin{acknowledgments}
This work was supported by a grant to JKT from the US-Norway Fulbright Foundation for Educational Exchange. AAK acknowledges funding from NSF Career Award \#1054671.
\end{acknowledgments}

\bibliography{./Tromborg}

\clearpage
\section{Supplemental Figure}
\renewcommand{\thefigure}{S\arabic{figure}}
\renewcommand{\thetable}{S\arabic{table}}
\renewcommand{\trianglelo}{\protect\marksymbol{triangle}{black}{90}{1.4}}
\renewcommand{\trianglelf}{\protect\marksymbol{triangle*}{black}{90}{0.8}}
\renewcommand{\diamondo}{\protect\marksymbol{diamond}{black}{0}{1.4}}
\renewcommand{\diamondf}{\protect\marksymbol{diamond*}{black}{0}{0.8}}
\renewcommand{\circleo}{\protect\marksymbol{o}{black}{0}{1}}
\renewcommand{\circlef}{\protect\marksymbol{*}{black}{0}{0.6}}
\renewcommand{\squareo}{\protect\marksymbol{square}{black}{0}{1}}
\renewcommand{\squaref}{\protect\marksymbol{square*}{black}{0}{0.6}}
\renewcommand{\triangledo}{\protect\marksymbol{triangle}{black}{180}{1.4}}
\renewcommand{\triangledf}{\protect\marksymbol{triangle*}{black}{180}{0.8}}

\begin{figure}
\centering
\includegraphics[width=\columnwidth]{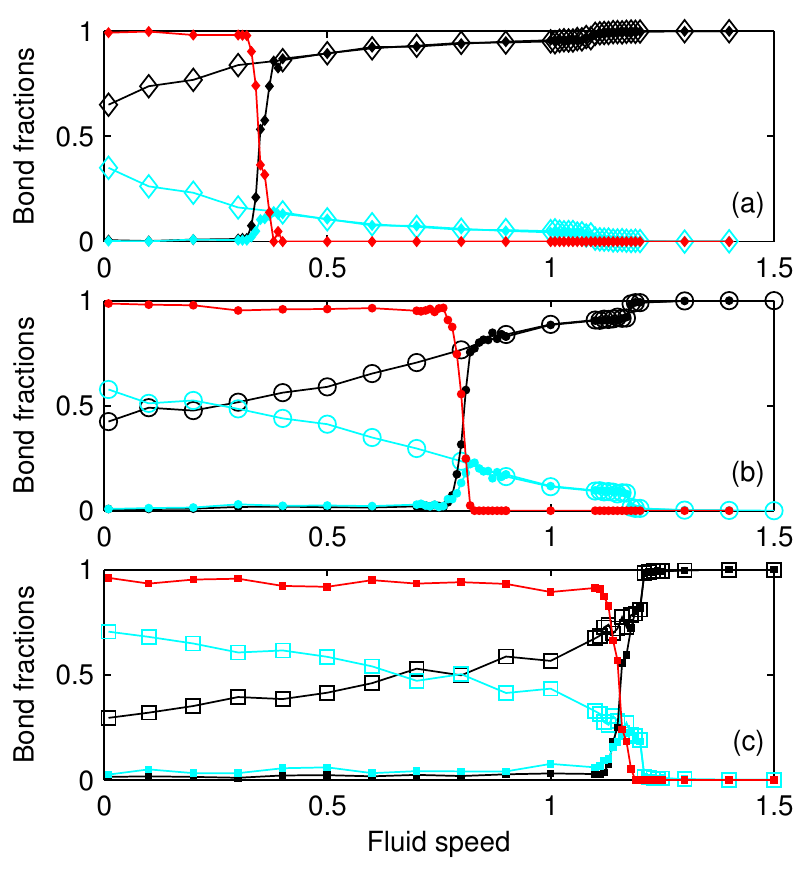}%
\caption{Average bond fractions reached in steady state. In the simulations with only weak bonds, the average fraction of formed bonds has an appreciable negative slope throughout the range where bonds are formed. In the simulations with strong bonds as well, the fraction of bonds formed stays nearly constant up to $\Vfluidtransition$, and then drops sharply. This is consistent with our observation in the main text that strong bonds arrest the object, but weak bonds do not, because the bond fraction depends on the slider speed, which becomes independent of $\Vfluid$ as long arrest occurs. Legend: fractions of unformed (\blackdash), weak (\cyandash) and strong (\reddash) bonds for $\Ntype=[1,\,0]$ (\diamondo), $[1,\,1]$ (\diamondf), $[3,\,0]$ (\circleo), $[3,\,1]$ (\circlef), $[5,\,0]$ (\squareo) and $[5,\,1]$ (\squaref).\label{fig:steady_state_bond_fractions}}
\end{figure}

\Fig~\ref{fig:steady_state_bond_fractions} shows the average bond fractions in the steady state reached at the end of our simulations. Each panel combines $\Ntype = [N_\text{weak},\,0]$ and $[N_\text{weak},\,1]$, with $N_\text{weak}=1$, $3$, $5$. We measured the average bond fractions in the last $10$ units of simulation time for the same simulations as in \fig~\ref{fig:Vslider_vs_Vfluid}. Where there are multiple simulations with identical settings, the average over simulations was taken as well.

\section{Supplemental Code}
Our computer code is available from arXiv as a separate download.

\end{document}